\documentclass[conference,letterpaper,10pt]{IEEEtran} %,compsoc
\usepackage{times,subfigure}
\usepackage{graphicx,algorithm,algorithmic,multirow,flushend}
\usepackage{latexsym,color,float,amsmath,amssymb}
\usepackage{pgf,tikz}
\usepackage{flushend}

\DeclareMathOperator{\Tr}{Tr}

\IEEEoverridecommandlockouts

\begin{document}

%latex2html -split 0 ancol-coop.tex

\title{Distributed Space-Time Coding of Over-the-Air Superimposed Packets in Wireless Networks\thanks{Antonios Argyriou would like to acknowledge the support from the European Commission through the Marie Curie Intra-European Fellowship WINIE-273041 and the STREP project CONECT
(FP7ICT257616).}}

\author{\IEEEauthorblockN{Antonios Argyriou\IEEEauthorrefmark{1}\IEEEauthorrefmark{2}
}
\IEEEauthorblockA{\IEEEauthorrefmark{1}Department of Computer and Communication Engineering, University of Thessaly, Volos, 38221, Greece}%
 \IEEEauthorblockA{\IEEEauthorrefmark{2}The Center for Research and Technology Hellas (CERTH), Thessaloniki, 57001, Greece%\\ %Email: anargyr@ieee.org
 }

}
\maketitle%

%\graphicspath{{figures/}}

\begin{abstract}
%Concurrent transmission and interference of wireless signals has been can be exploited for incdrereasing the system throughout through multi-user detection, multiple packet reception and physical layer network coding.
In this paper we propose a new cooperative packet transmission scheme that allows independent sources to superimpose over-the-air their packet transmissions. Relay nodes are used and cooperative diversity is combined with distributed space-time block coding (STBC). With the proposed scheme the participating relays create a ST code for the over-the-air superimposed symbols that are received locally and without proceeding to any decoding step beforehand. The advantage of the proposed scheme is that communication is completed in fewer transmission slots because of the concurrent packet transmissions, while the diversity benefit from the use of the STBC results in higher decoding performance. The proposed scheme does not depend on the STBC that is applied at the relays. Simulation results reveal significant throughput benefits even in the low SNR regime.

\end{abstract}

%\begin{IEEEkeywords}
%Cooperative protocol, Alamouti code, space-time block code, distributed space-time code, physical layer network coding, superposition modulation, superimposed signals, wireless networks.
%\end{IEEEkeywords}

\section{Introduction}
Improving the reliability and throughput of wireless services has always been one of the main motivations of the wireless communications research community. Towards these two objectives, the recent years there is a significant interest around the idea of allowing packets to interfere or be superimposed over the air. The term physical layer network coding (PLNC) is also used for the previous concept. With PLNC, interference is used constructively for maximizing system throughput and in most cases of PLNC-based systems it is something that is controlled. To be effective, there is a need for a type of node cooperation. Unlike the first works that considered this form of cooperation with the mindset towards maximizing the throughput of point-to-point links with bidirectional traffic~\cite{dankberg97,zhang:physical-layer-nc,katabi07a}, PLNC can have more substantial impact on the performance of more complex and even non-canonical wireless networks~\cite{avestimehr11}. The general idea in all these works is that wireless signals are allowed to interfere or to be superimposed. The problems that arise during the signal recovery process can either be addressed with pre-coding steps~\cite{ding11}, decoding at the relay if possible~\cite{akino09}, use of a-priori knowledge at the receivers~\cite{dankberg97,katabi07a}, and specialized decoder design~\cite{akino09,argyriou:twc-ancol}. One important result from the aforementioned works is that the existence of a-priori knowledge in the form of already decoded packets, is the only way to increase the decoding performance if more than two signals interfere~\cite{katabi07a,argyriou:twc-ancol}. Furthermore, all the aforementioned schemes do not process the mixed signal at the relays for further performance improvements.

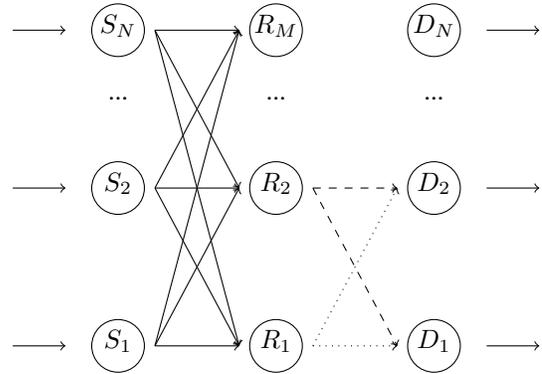
\begin{figure}[t]
\centering
\begin{tikzpicture}[scale=0.7]
  %\draw (0,0) -- (1.5,0) -- (1.5,1) -- (0,1);
  \draw (0.0,0.5) circle(0.5cm);
  \draw (0.0,3.5) circle(0.5cm);
  \draw (0.0,6.5) circle(0.5cm);
  \draw (0.0,0.2) node[above]{$S_1$} ;
  \draw (0.0,3.2) node[above]{$S_2$} ;
  \draw (0.0,6.2) node[above]{$S_N$} ;

    \draw (6.0,5.0) node[above]{...} ;
    \draw (3.0,5.0) node[above]{...} ;
    \draw (0.0,5.0) node[above]{...} ;

  \draw (3.0,0.5) circle(0.5cm);
  \draw (3.0,3.5) circle(0.5cm);
  \draw (3.0,6.5) circle(0.5cm);
  \draw (3.0,3.2) node[above]{$R_2$} ;
  \draw (3.0,0.2) node[above]{$R_1$} ;
  \draw (3.0,6.2) node[above]{$R_M$} ;

  \draw (6.0,0.5) circle(0.5cm);
  \draw (6.0,3.5) circle(0.5cm);
  \draw (6.0,6.5) circle(0.5cm);
  \draw (6.0,3.2) node[above]{$D_2$} ;
  \draw (6.0,0.2) node[above]{$D_1$} ;
  \draw (6.0,6.2) node[above]{$D_N$} ;

  \draw[->] (-2,3.5) -- node[above]{} (-1,3.5);
  \draw[->] (-2,0.5) -- node[above]{} (-1,0.5);
  \draw[->] (-2,6.5) -- node[above]{} (-1,6.5);
  \draw[->] (7,3.5) -- node[above]{} (8,3.5);
  \draw[->] (7,0.5) -- node[above]{} (8,0.5);
  \draw[->] (7,6.5) -- node[above]{} (8,6.5);

  \draw[->] (0.7,3.5) -- node[above]{} (2.3,3.5);
  \draw[->] (0.7,0.5) -- node[above]{} (2.3,0.5);
  \draw[->] (0.7,6.5) -- node[above]{} (2.3,6.5);

  \draw[->] (0.7,3.5) -- node[below right]{} (2.3,0.5);
  \draw[->] (0.7,0.5) -- node[above right]{} (2.3,3.5);
  \draw[->] (0.7,6.5) -- node[above right]{} (2.3,3.5);
  \draw[->] (0.7,3.5) -- node[below right]{} (2.3,6.5);

  \draw[->] (0.7,0.5) -- node[below right]{} (2.3,6.5);
  \draw[->] (0.7,6.5) -- node[below right]{} (2.3,0.5);

  \draw[->,dashed] (3.7,3.5) -- node[above]{} (5.3,3.5);
  \draw[->,dotted] (3.7,0.5) -- node[above]{} (5.3,0.5);

  \draw[->,dashed] (3.7,3.5) -- node[below right]{} (5.3,0.5);
  \draw[->,dotted] (3.7,0.5) -- node[above right]{} (5.3,3.5);

\end{tikzpicture}
\caption{System model for the cooperative scheme that employs distributed \emph{Space-Time Superimposed Symbol Coding (STSSC)}. Different line styles indicate transmissions in different time slots. For the channels with the dashed lines not all connections are depicted to avoid clogging the figure.
}
\label{fig:model_sr}
\end{figure}

%"In this paper, we present a novel high-throughput medium
%access scheme for wireless networks that exploits the broad-
%cast nature of the wireless channel and the virtual arrays pro-
%vided by users in the system who are willing to cooperate. Mul-
%tiple transmissions over the same slot are not treated as colli-
%sions.Instead,once a multiple transmission has been detected, a
%set of nodes designated as non-regenerative relays forward what
%they received during the collision slot one by one in a predeter-
%mined order. By processing the originally transmitted packets
%and those forwarded by the relays, the destination node can re-
%cover the original packets. Our goal is to develop a scheme that
%maintains the benefits of ALOHA systems in the sense that all
%nodes share access to media resources efficiently and without
%extra scheduling overhead, and at the same time provides the
%benefits of multi-antenna systems, e.g., diversity, without addi-
%tional antenna hardware."

%STC is undoubtedly one major leap forward in wireless communications. The original works on multi-antenna STC~\cite{alamouti98,tarokh99}, were followed in the recent years from a significant amount of efforts that extend STC schemes from multi-antenna systems to relay-based cooperative systems.
A different class of works that employs relay processing, but on non-interfering signals, is the randomized distributed space time coding (R-DSTC)~\cite{bouyer01}. With this scheme sources transmit one at a time, and then the relay nodes forward  a random linear combination of the decoded packets. One disadvantage of systems that fully decode the signals at the relay is the low achieved rate. %The issue is that the number of relays that can decode the transmitted message is increased if the transmission rate is being reduced. This approach, however, will require a substantial reduction of the rate.
To alleviate this issue, the authors in~\cite{lai11} developed space-time network coding. The notion of space-time cooperation in that work comes from the use of multiple relays and algebraic network coding across different packet transmissions. A similar idea that requires no decoding at the relays, but at the same time it does not exploit any form of interfering signals, was presented in~\cite{jing06}. In that work the main idea is that the transmitter sends a packet in one slot, while in a number of subsequent slots the relays encode their received signals into a "distributed" linear dispersion (LD) code, and then they transmit the coded signals to the destination node.

%The general idea of the distributed STCs is that "Cooperative schemes in general contain two phases of trans- mission, namely the broadcast phase and the cooperation phase. During the cooperation phase, relays collaboratively transmit the re- encoded source information, where ’virtual’ space-time codewords can be formed." On the other hand, one of the disadvantage of our previous scheme was that it employed a type of distributed spatial multiplexing. While Spatial Multiplexing achieves high data rates, it does not succeed in leveraging transmit diversity.

In our effort to increase the throughput performance of wireless networks, and allow more nodes to superimpose their signals over-the-air, in this paper we propose a new form of distributed space-time-coded (STC) cooperation that is named distributed \textit{space-time superimposed symbol coding (STSSC)}. Our scheme aims at a generalization of distributed STCs for the case of signals that are mixed or superimposed over-the-air. The STC is created by linearly coding across time different versions of the same superimposed symbols. %The key difference from related works is that the different space-time coded versions of the same superimposed signals improve the decoding performance of individual symbols and eventually improve the performance of PLNC.
% At the STC decoder the different versions of the superimposed signals that are obtained are used with a maximum-likelihood (ML) decoder. Another key advantage is that the proposed scheme is independent of the type of the employed STC. Finally it must be noted that the cooperative scheme exploits the concurrent transmission and reception of packets from several independent unicast senders, while several nodes can act as relays.

\section{System Model and Overview}
\label{section:system-model}
Our study focuses on the relay network model where a group $\mathcal{S} = \{S_1,S_2,...,S_N\}$ of sources want to communicate with a group $\mathcal{D} = \{D_1,D_2,...,D_N\}$ of destinations with the assistance of a set $\mathcal{R} = \{R_1,R_2,...,R_M\}$ relays. Every node has a single omnidirectional antenna that can be used both for transmission and reception and they all have the same average power constraint. We denote the channel from the $s$-th transmitter to the $r$-th relay as $h_{s,r}$, and the channel from the $r$-th relay to destination $d$ as $h_{r,d}$. We also assume that $h_{s,r}$ and $h_{r,d}$ are independent complex Gaussian random variables with zero-mean and unit-variance. In Fig.~\ref{fig:model_sr} we present the topology that we study in this paper and it includes the sources, the relays, and the destinations. The transmission of a packet requires two hops since we assume that there is no direct link between the sources and the destinations. All the channels, from sources to relays and relays to destinations are considered to be block-fading Rayleigh. The channel is assumed to remain constant for the coherence period of the channel that is $T$ symbols and each source/relay and relay/destination pair. Additive white Gaussian noise (AWGN) with zero mean and unit variance is assumed at the relays and the destinations.

\begin{figure}[t]
\begin{center}
%\subfigure[Distributed Space-Time Coding]{ \includegraphics[keepaspectratio,width = 0.99\linewidth]{figures/dispersion_code_basic.eps}}
%\subfigure[Distributed Space-Time Superimposed Symbol Coding]{ 
\includegraphics[keepaspectratio,width = 0.99\linewidth]{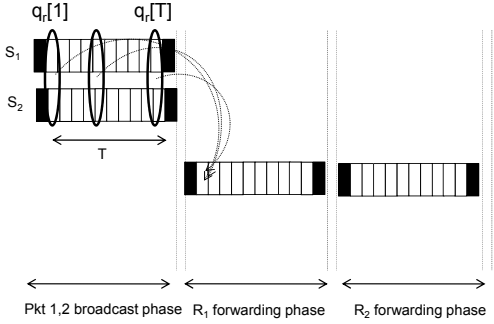}
%}
\end{center}
  \caption{Behavior of STSSC in the time domain. With STSSC the signal that is forwarded during one symbol slot is a linear combination (dashed lines) of all the superimposed symbols $q_r$ that were received during the broadcast phase (grey ellipses).}
 \label{fig:dispersion_code}
\end{figure}

%The input at each source is a sequence of $B$ bits and its output is a sequence of $U$ symbols. This sequence of symbols is transmitted and superimposed over-the-air at each relay.
To model the system more effectively we introduce the term \emph{symbol slot} which is the basic time unit that we consider in this paper and it corresponds to the transmission time of a PHY symbol. A \emph{transmission phase} is a system defined parameter and denotes the time period where a user (source or relay) can transmit a packet and it consists of many symbol slots. %This is essentially similar to a time-division multiple access (TDMA) slot but in our system multiple users can transmit at the same slot. At the relays the non-decodable superimposed symbol is then spread in space (over the $R$ relays) and in time (over $T$ symbol durations) in order to form a codeword.
%represented by a matrix $C$ of size $R \times T$. The ratio $N\times B/T$ is the signaling rate of the transmission, whereas the ratio $R\times U/T$ is defined as the spatial multiplexing rate of the space-time code. The latter is representative of how many symbols are packed within a codeword per unit of time which depends on the used code.
%
%
The cooperative packet transmission requires $N$ transmission phases as defined previously. In the first  transmission phase, all the source stations transmit/broadcast the packet to all the potential relay stations. The relays do not try to decode but instead they apply a distributed STC to the signals that are received in a superimposed form locally, and then they forward them to the destinations. Note the key difference with existing distributed STC schemes is that information symbols are not available at the relay for decoding but only their superimposed versions. The superimposed signals can be coded with any type of distributed STC (e.g. orthogonal or LD) code, while then they are sequentially forwarded from the relays. This relay transmission order can be random but it should be decided in advance of a communication phase. At each destination the forwarded signals from the relays are jointly ML decoded while we assume receiver channel state information (CSI) is available.

\section{The distributed STSSC Scheme}
\label{section:analysis}
In this section we proceed by analyzing in detail the behavior of the proposed cooperative scheme. But first we define the paper notation. Matrices are denoted with bold capital letters, i.e. $\mathbf{A}$. Bold lowercase denote vectors. The matrices $\mathbf{A}^T$, $\mathbf{A}^H$,$\mathbf{A}^*$, are the transpose, Hermitian, and conjugate of $\mathbf{A}$. The Euclidian norm is denoted by $\|\cdot\|$ and $\Tr(\cdot)$ is the trace of a matrix.
%In the general case of the proposed scheme during the broadcast phase all the senders are allowed to transmit by being synchronized only at the symbol level. In that case the symbols would be superimposed in the time domain in an arbitrary fashion.

\subsection{Source to Relay Transmission Phase}
Let $c_{s}[t]$ denote the symbol that source $s$ wants to transmit during the symbol slot $t$. Then, the transmitted signal from all the sources is normalized to have unit energy during the coherence period of $T$ symbols.
%\begin{equation}
%E[\mathbf{X}_{\mathcal{S}}\mathbf{X}_{\mathcal{S}}^*]=T\mathbf{I}_T
%\end{equation}
The array of transmitted symbols is expressed as
\begin{eqnarray}
\mathbf{X}_{\mathcal{S}} =\frac{1}{\sqrt{TN}}\left [ \begin{array}{ccccc}
  c_{1}[1] & ... & c_1[t] & ... & c_1[T]\\
%  x_{2}[1] & ... & x_2[t] & ... & x_2[T]\\
  ... 	 & ... & .... & ... & ...\\
  c_s[1]    & ... & c_s[t] & ... & c_s[T]\\
  ... 	 & ... & ... & ... & ...\\
  c_{N}[1]& ... & c_N[t] & ... & c_N[T]
  \end{array}\right]. \nonumber
\end{eqnarray}
%OLD
%\begin{eqnarray}
%\mathbf{x}_{\mathcal{S}} = \left  [ \begin{array}{llll}
%  x_1[1]~x_2[1]~x_3[1]~...~x_N[1]~...~x_1[t]~x_2[t]~x_3[t]~...~x_N[t]~... x_1[T]~x_2[T]~x_3[T]~...~x_N[T]
%  \end{array}\right]^T \in\mathbb{C}^{NT\times 1},\nonumber%
%\end{eqnarray}
The energy normalization process in this case corresponds to $\Tr (E[\mathbf{X}_{\mathcal{S}}^H\mathbf{X}_{\mathcal{S}}])=1$, in order to compare fairly systems with different $N$ and $T$.
%For the complete period $T$ the matrix that is used for the transmitted symbols is the two-dimensional $T\times T$ block diagonal matrix $\mathbf{X}_{\mathcal{S}}=\text{diag}(\mathbf{x}_{\mathcal{S}}~\mathbf{x}_{\mathcal{S}}~\mathbf{x}_{\mathcal{S}}~...)\in \mathbb{C}^{NT^2\times T}$.

During the first communication phase that is described here, and from symbol slot 1 to $T$, a source $s\in\mathcal{S}$ transmits the signal $\sqrt{\rho}x_{s}[t]$ that is a part of larger packet. We assume that $E[|h_{s,r}|^2]$ = 1, and since the average signal power is normalized to unity, and the noise term has zero mean and 1/2 variance per dimension, $\rho$ can be interpreted as the average signal-to-noise ratio (SNR) at the receiver. With $w_r[t]$ we denote the sample of the AWGN during symbol slot $t$, and $\mathbf{h}_{\mathcal{S},r}\in \mathbb{C}^{1\times N}$ is the channel matrix that contains the $N$ channel gains from the first broadcast phase, i.e. from the group of sources $\mathcal{S}$ to relay $r$:
\[
 \mathbf{h}_{\mathcal{S},r} = [  h_{1,r}~~h_{2,r}~...~h_{N,r}]
\]
Now we can write in a compact form and with a vector notation the received signal during one symbol slot as
\[
q_r[t] =\sqrt{\rho} \mathbf{h}_{\mathcal{S},r} \mathbf{x}_{\mathcal{S}}[t]+w_r[t],
\]
%\begin{equation}\label{eqn:q_r}
%\end{equation}
where $\mathbf{x}_{\mathcal{S}}[t]\in \mathbb{C}^{N\times 1}$ is a column vector of the matrix $\mathbf{X}_{\mathcal{S}}$. Note that at this stage the relay does not have a decodable signal\footnote{The relay can decode with ML or an efficient sphere decoder but still the high BER makes this approach impractical.} and so it cannot construct a STC for a specific symbol $x_s[t]$. Instead, the relay constructs a STC for a superimposed signal that is a linear combination of several symbols.

\subsection{Relay Operation}
%\subsection{Relay Operation under Protocol I - STC for Superimposed Symbols}
%The most important functionality of the proposed system is exercised at the relays. With the baseline PLNC only amplify-and-forward (AF) is exercised by the relays after the symbols are locally superimposed~\cite{zhang:physical-layer-nc,katabi07a,argyriou:twc-ancol}. If the AF scheme is employed during the forwarding phase from relay $r$, the received signal at $d$ can be written as follows:
%\begin{equation}
%\label{eqn:y_Dk_Rm}
%y_{r,d}[t'] = g_r h_{r,d}q_r[t]+ w_d[t']
%\end{equation}
%In the previous equation $g_r$ is the power scaling factor that is applied by the relay, $q_r[t]$ is one element of the matrix $\mathbf{q}_r$, while $w_d[t']$ is the AWGN sample at the destination. In this paper the previous PLNC scheme is named \emph{amplify-and-forward of over-the-air superimposed transmissions} (AF-OST)~\cite{argyriou:twc-ancol} and it will be used for performance comparison with STSSC.
%If all $R$ relays broadcast at the same time:
%\begin{eqnarray}\label{eqn:y_Dk_Rm}
%y_{k,\mathcal{R}}[t'] &=& \sum_{r\in\mathcal{R}}  h_{r,d} \cdot (g_r[t] \cdot q_r[t])+ w_d.
%\label{fp1}
%\end{eqnarray}
%\textbf{If a subset of the relays??}
%Now let us continue with the description of the main idea of this paper.
With STSSC we design the transmit signal at every relay $r$ as a linear function of its received non-decodable over-the-air superimposed symbol (OSS) $q_r$. This is done for each symbol while there are in total $T$ symbols. %In this case, for \textbf{one} symbol that will be forwarded in symbol slot $t'$:
%\begin{eqnarray}\label{eqn:z_r1}
%z_{r}[t'] &=& \sum^{T}_{t=1} g_r[t] \alpha_{r,t' t}q_r[t]
%\end{eqnarray}
If an orthogonal STBC is used, or even a more general linear dispersion (LD) code, the transmitted signal will be described in the following $T\times T$ matrix:
\begin{equation}\label{eqn:z_r2}
\mathbf{Z}_r=g_r \sum_{t=1}^T(\mathbf{A}_t q_r[t]+\mathbf{B}_t q^*_r[t])
\end{equation}
Each column of this matrix is transmitted simultaneously from the relay $r$. All the $M$ relays  transmit similarly in their corresponding transmission phase. $\mathbf{A}_t,\mathbf{B}_t$ are the STC matrices~\cite{book:coding-for-mimo}. What this expression demonstrates, is that a symbol to be transmitted in a forwarding symbol slot, is a linear combination of all the received symbols in the previous $T$ symbol slots. %The notation $\alpha_{t t'}$ denotes the coefficient from the selected code that is multiplied with the received signal during the broadcast phase and symbol slot $t$, for generating the signal that will be transmitted from relay $r$ in the forwarding phase and during symbol slot $t'$. During the same forwarding phase and another symbol slot $t''$ then a similarly formed signal will be transmitted but in that case different coefficients $\alpha$ and $\beta$ will be used depending on the used STC.
This process is also clearly visible in Fig.~\ref{fig:dispersion_code} where the creation process of the ST-coded superimposed symbols is depicted graphically. %Since each relay constructs similarly the signal $z$, the scheme constitutes a distributed STC of over-the-air superimposed symbols.

Let us elaborate the previous expression and re-write it as:
\begin{eqnarray*}
\mathbf{Z}_r&=& \sqrt{\rho} g_r \sum_{t=1}^T \sum_{s=1}^N   ( h_{s,r} \mathbf{A}_t x_s[t]+  h^*_{s,r} \mathbf{B}_t  x^*_s[t])\\
&+&g_r \sum_{t=1}^T  (\mathbf{A}_t w_r[t]+\mathbf{B}_t w^*_r[t])
\end{eqnarray*}
%In the above $\mathbf{h}_{\mathcal{S},r}\mathbf{x}_{\mathcal{S}}[t]$ is scalar.
At the destination, the received signal from one relay will then be
\begin{eqnarray}\label{eqn:y_rd}
\mathbf{y}_{r, d} &=&\sqrt{\rho}   g_r  \mathbf{h}_{r,d} (\sum_{t=1}^T  \sum_{s=1}^N ( \mathbf{A}_t h_{s,r}x_{s}[t]+  \mathbf{B}_t h^*_{s,r}x^*_{s}[t])\nonumber\\
&+&\sum_{t=1}^T  (\mathbf{A}_t w_r[t]+\mathbf{B}_t w^*_r[t])) +\mathbf{w}_{r,d},
\end{eqnarray}
where $\mathbf{h}_{r,d}\in \mathbb{C}^{1\times T}$ contains the channel gain during $T$ symbol slots and it remains unchanged according to our stated assumptions. The previous expression is important since it demonstrates that with this formulation and system design, we are able to express the signal at the receiver as a function of the transmitted signal from the sources and not just that of the superimposed signal at the relay. Thus, for each OSS that will be transmitted, a new STC is created by forming a linear combination of all the symbols received in the previous transmission phase that has a duration of $T$ symbols.
%In \eqref{eqn:y_Dk_r} $\mathbf{A}_t \mathbf{h}_{\mathcal{S},r}\mathbf{X}_{\mathcal{S}}[t]+\mathbf{B}\mathbf{h}^*_{\mathcal{S},r}\mathbf{X}^*_{\mathcal{S}}[t]$ is the space-time code. This matrix is equivalent to the space-time code in a multiple-antenna or a distributed cooperative system.
The proposed scheme can also be classified as distributed STSSC since the ST code is applied in a distributed fashion by the relays, while it is also based in ST coding of superimposed symbols.

%\textbf{TBD:With the concurrent TX protocol it will be }
%\begin{eqnarray*}
%\mathbf{y}_{d} =\sum^M_{r=1} \mathbf{y}_{r,d}=\left [ \begin{array}{cc}
%  \quad \sum^M_{r=1}  h_{r,d} (q_r[1]-q^*_r[2]) & \sum^M_{r=1}  h_{r,d}( q_r[2]+q^*_r[1])
%  \end{array}\right],
%\end{eqnarray*}

%After the broadcast phase is completed, and after the signal $z$ is created at each relay, there are multiple forwarding phases, i.e. their precise number is equal to the number of transmitting sources $N$. In each of the forwarding phases a relay $r$ broadcasts the received signals that are transmitted as specified in~\eqref{eqn:z_r2}
The relay also applies power scaling so as to maintain the power constraint. If $\sigma^2$ is the noise variance, then the power scaling is given as
\begin{eqnarray}\label{eqn:power-constraint1}
g_r =  \sqrt{\frac{\rho}{\rho\sum^{N}_{s=1} | h_{s,r}|^2 +\sigma^2}}.
\end{eqnarray}
We can express in a more concise form the matrix of the power amplification for all relays as the following $M\times M$ matrix:
\begin{eqnarray}
\mathbf{G} =  \text{diag} (  g_1 \quad ... \quad  g_r \quad ... \quad g_{M} )
\end{eqnarray}

\section{Decoding}
\label{section:decoding}
The proposed scheme can work with an arbitrary LD code. However, for keeping the analysis simple and for demonstrating the main concept of the decoding approach more clearly, we assume that the ST code is orthogonal. Based on this choice, we present a new decoding algorithm that combines the classic approach for decoding general orthogonal designs and ML decoding for decoding superimposed/interfering symbols~\cite{book:coding-for-mimo}.
To proceed with the description of the decoding process let us first define an extended form of the signal that is received during one forwarding phase from relay $r$ and its complex conjugate as follows
\begin{equation}\label{eqn:y_rd_conjugate}
\mathbf{\tilde{y}}_{r,d} = [
  y_{r,d}[1] \quad   ... \quad y_{r,d}[T] \quad y^*_{r,d}[1] ... \quad y^*_{r,d}[T]
]
\end{equation}
%At this point we have to clarify that decoding is performed for calculating the symbols that were transmitted during the symbol slot $t'$ that are denoted as $\mathbf{\hat{X}}_{\mathcal{S}}[t']$.
The primary ML decision problem we desire to solve is equivalent to minimizing the squared Euclidean distance metric~\cite{book:coding-for-mimo}. From~\eqref{eqn:y_rd_conjugate} and~\eqref{eqn:y_rd} we have that this ML metric is
\begin{eqnarray}
\label{eq:ml-decoding1}
e &=& \sum^{M}_{r=1} \Big \| \mathbf{\tilde{y}}_{r, d}
- g_r  \sqrt{\rho} \sum^{T}_{t=1} \sum^{N}_{s=1} \Big (   \left [ \begin{array}{cc} \mathbf{h}_{r,d}\mathbf{A}_{t} h_{s,r}\\ \mathbf{h}^*_{r,d}\mathbf{B}^*_t h_{s,r}  \end{array}\right]^T x_{s}[t]\nonumber\\
&+& \Big [  \mathbf{h}_{r,d}\mathbf{B}_t h^*_{s,r} \quad \mathbf{h}^*_{r,d}\mathbf{A}^*_t h^*_{s,r}  \Big ]  x^*_{s}[t]\Big )\Big \|^2
\end{eqnarray}
The decoding expression of~\eqref{eq:ml-decoding1} can be simplified to
\begin{eqnarray}
\label{eq:ml-decoding3}
e&=& \| \mathbf{\tilde{y}}_{r,d} \|^2 -2g_r \sqrt{\rho} \Tr \Big ( \left [ \begin{array}{cc} \mathbf{h}_{r,d}\mathbf{A}_{t} h_{s,r}\\ \mathbf{h}^*_{r,d}\mathbf{B}^*_t h_{s,r}  \end{array}\right]^* \mathbf{\tilde{y}}^H_{r,d}  x_s[t]\Big )\nonumber\\
&+&g^2_r \rho|h_{s,r}|^2 \|\mathbf{h}_{r,d}\|^2 \Tr (\mathbf{A}^H_{t}\mathbf{A}_{t}+\mathbf{B}^H_{t}\mathbf{B}_{t} ) | x_s[t] |^2
\end{eqnarray}
At this stage, one of the key ideas of the decoding algorithm is to employ matched filtering for each symbol that was transmitted at each source. Thus, we have that the sufficient statistic we can get for each symbol is
\begin{eqnarray*}
\label{eq:ml-decoding4}
u_{s,t}= \Tr \Big ( \sum^{M}_{r=1} g_r \Big [ \mathbf{h}_{r,d}\mathbf{A}_{t} h_{s,r} \quad \mathbf{h}^*_{r,d}\mathbf{B}^*_t h_{s,r}  \Big ]^H \mathbf{\tilde{y}}_{r, d} \Big ) %\\
%&=&\sum^{M}_{r=1}  h^*_{s,r}g_r  \Big ( \mathbf{A}^H_{t}\mathbf{h}^H_{r,d} \mathbf{y}_{r,d}+\mathbf{y}^H_{r,d}\mathbf{h}_{r,d}\mathbf{B}_{t} \Big ).
\end{eqnarray*}
Due to the properties of orthogonal STBCs we have that
\begin{eqnarray*}
\label{eq:ml-decoding6}
u_{s,t}&=&\sum^{M}_{r=1}  h^*_{s,r}g^2_r \sqrt{\rho} \| \mathbf{h}_{r,d}\|^2  \Tr \Big ( \mathbf{A}^H_{t}\mathbf{A}_t+\mathbf{B}^H_t\mathbf{B}_{t} \Big )\\
&\times & \sum_{n=1}^N h_{n,r}x_n[t] +\sum^{M}_{r=1}  h^*_{s,r}g^2_r \sqrt{\rho} \| \mathbf{h}_{r,d}\|^2   \Tr  \Big (  ( \mathbf{A}^H_{t}\mathbf{A}_t\\ &+&\mathbf{B}^H_t\mathbf{B}_{t} ) w_r[l] +\mathbf{A}^H_{t}\mathbf{h}^H_{r,d}\mathbf{w}_{r,d} +\mathbf{w}^H_{r,d}\mathbf{h}_{r,d}\mathbf{B}_{t}\Big ).
\end{eqnarray*}
This last expression is important since it demonstrates clearly that matched filtering for a symbol transmitted during slot $t$, decouples all the other symbols that we transmitted during any of the remaining symbol slots. Thus, the decoder processing is linear with respect to the number of $T$ symbols transmitted from each source (for each symbol slot $t$ a new group of symbols is decoded) but not with respect to the number of sources.

%\subsection{Final ML decoding step}
After obtaining the sufficient statistic $u_{s,t}$  for each specific information symbol $x_{s,t}$, decoding cannot proceed based only on this information. Without interference/superimposed symbols, this would normally be the case. Now, the algorithm must account the fact that $u_{s,t}$ contains the impact of all the other symbols that were transmitted in the specific symbol slot $t$. To proceed with the final steps of the decoding algorithm we define the following scalar quantity in~\eqref{eq:ml-decoding3} :
\begin{eqnarray}
\label{eq:ml-decoding4}
v_{s,t}= \sum^{M}_{r=1}  g^2_r d_{s,t} | h_{s,r} |^2   \| \mathbf{h}_{r,d} \|^2
\end{eqnarray}
%Then the final metric is:
%\begin{eqnarray}
%\label{eq:ml-decoding2}
%e=\sum^{M}_{r=1}  \| \mathbf{\tilde{y}}_{r,d} \|^2-2 \sqrt{\rho}\sum^{T}_{t=1} \sum^{N}_{s=1} u_{s,t}x_s[t] + \rho\sum^{T}_{t=1}\sum^{N}_{s=1} v_{s,t}| x_s[t] |^2
%\end{eqnarray}
The decoding proceeds as follows. The sufficient statistic for each symbol $s,t$ is calculated with matched filtering, and then the metric for each symbol is obtained from~\eqref{eq:ml-decoding3} as
\begin{eqnarray}
\label{eq:ml-decoding2}
e_{s,t}=\sum^{M}_{r=1}   \| \mathbf{\tilde{y}}_{r,d} \|^2-2 \sqrt{\rho} u_{s,t}x_s[t] +  \rho v_{s,t}| x_s[t] |^2.
\end{eqnarray}
To decode the concurrently transmitted symbols we employ ML decoding. We write the ML rule for all sources as follows:
\begin{eqnarray}
\label{eq:ml-decoding2}
\mathbf{\hat{x}}_{S}[t]&=& \arg \min_{\mathbf{x}_{S}[t]\in \mathbb{C}^N}  \sum_{s=1}^{N} e_{s,t}
\end{eqnarray}
Therefore, each group of symbols that are transmitted concurrently in symbol slot $t$ is simultaneously decoded with ML decoding and the estimated result is the vector $\mathbf{\hat{x}}_{\mathcal{S}}[t]$. This is another important objective of this algorithm, i.e. to increase the diversity of interfered/superimposed symbols by introducing ST coding and linear processing.

%To finalize the description of the decoding algorithm we provide expressions for the Alamouti-based example. The sufficient statistics for the symbols that are transmitted during the first symbol slot, and for one relay in the Alamouti example are
%\begin{eqnarray*}
%u_{1,1}&=&h^*_{1,r}g_r \Tr \Big ( \mathbf{A}^H_{1}\mathbf{h}^H_{1,d} \mathbf{y}_{1,d}+\mathbf{y}^H_{1,d}\mathbf{h}_{1,d}\mathbf{B}_{1} \Big )\\
%&=&h^*_{1,r}g_r \Tr \Big ( \mathbf{A}^H_{1} | h_{r,d}|^2 \left [ \begin{array}{cc}
%  q_r[1] & q_r[2] \\
%  -q^*_r[2] & q^*_r[1]
%  \end{array}\right]\\
%  &+&\left [ \begin{array}{cc}
%  q^*_r[1] & -q_r[2] \\
%  q^*_r[2] & q_r[1]
%  \end{array}\right]| h_{r,d}|^2\mathbf{B}_{1} \Big )\\
%  &=&h^*_{1,r}g_r| h_{r,d}|^2 (q_r[1]+q_r[1])\\
%  &=&h^*_{1,r}g_r| h_{r,d}|^2 2(h_{1,r}x_1[1]+h_{2,r}x_2[1]+w_1)\\
%  u_{2,1}&=&h^*_{2,r}g_r| h_{r,d}|^2 2(h_{1,r}x_1[1]+h_{2,r}x_2[1]+w_1)
%\end{eqnarray*}

\section{Performance Evaluation}
\label{section:performance-evaluation} We implemented the proposed cooperative scheme and we evaluated the performance in terms of BER and throughput under different channel conditions through Monte Carlo simulations. Simulation results are presented for the point-to-point transmission mode (named $Direct$ in the figures) as a reference for all our results. We also implemented the \textit{Distributed STC} protocol, where transmissions occur independently without being superimposed and are subsequently decoded at the relay. The \emph{Distributed STC} is then applied at the relays and the transmitted signals are combined with maximum ratio combining (MRC) and ML decoding at each destination~\cite{argyriou:twc-ancol}. Finally we also evaluated the performance of a scheme named $AF-OST$ where packets are superimposed in the broadcast phase while they are sequentially amplified and forwarded for subsequent ML decoding at the destination. We present the averaged results for 2000 packet transmissions that have a packet length of 1000 bits. The channel bandwidth is 20 MHz, while the same path loss model was used for all the channels. Furthermore, we also assume that the noise over the wireless spectrum is AWGN with the variance of the noise to be $10^{-9}$ W/Hz at every node/link. We also used a Rayleigh fading wireless channel model. The channel transfer functions between the nodes vary independently but they are characterized by the same average SNR unless otherwise specified. Therefore, the BER of one of the destinations is only plotted.

\begin{figure}[t]
\begin{center}
\subfigure[BER]{ \includegraphics[keepaspectratio,width = 0.495\linewidth]{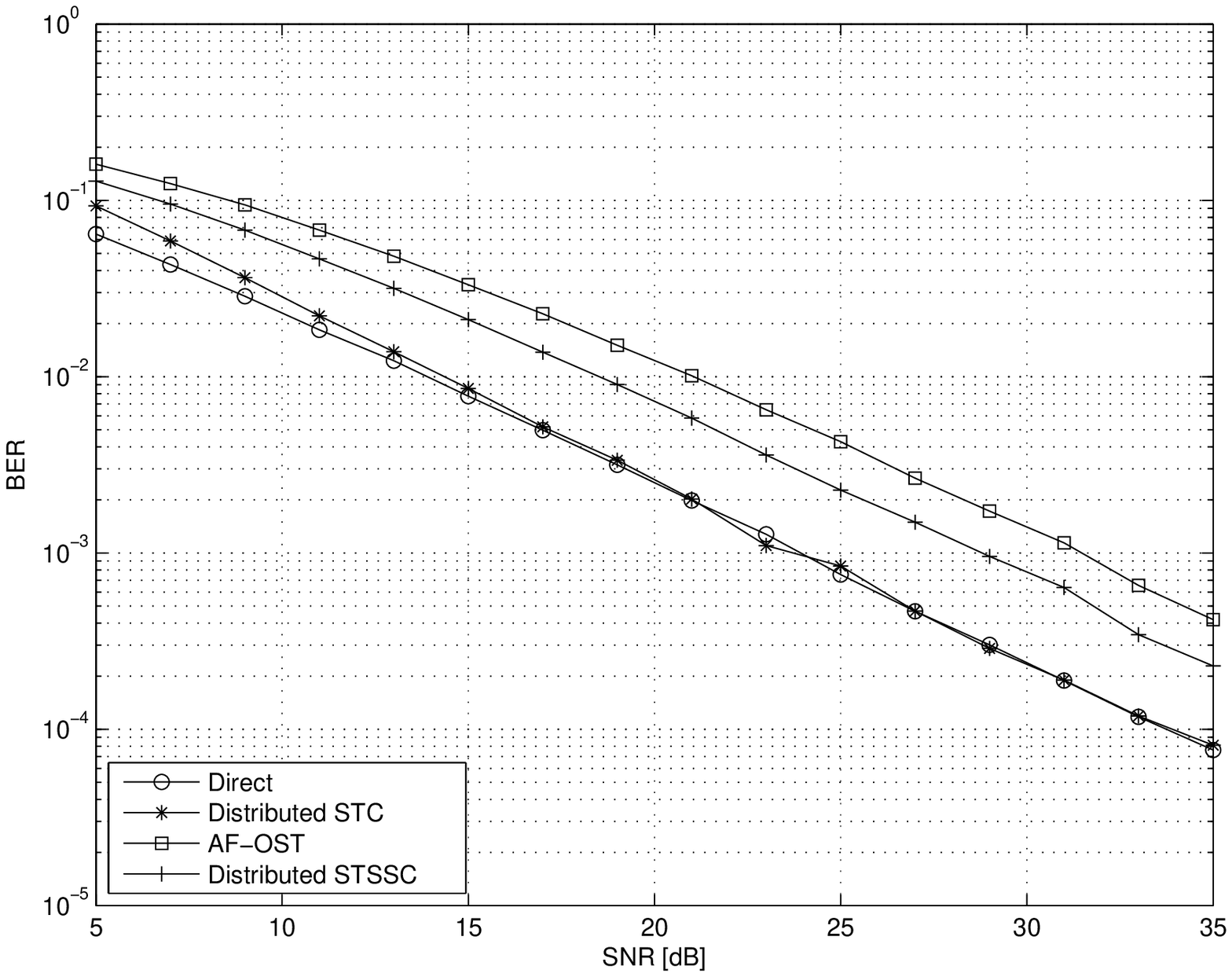}}\hspace{-0.3cm}
\subfigure[Throughput]{ \includegraphics[keepaspectratio,width = 0.495\linewidth]{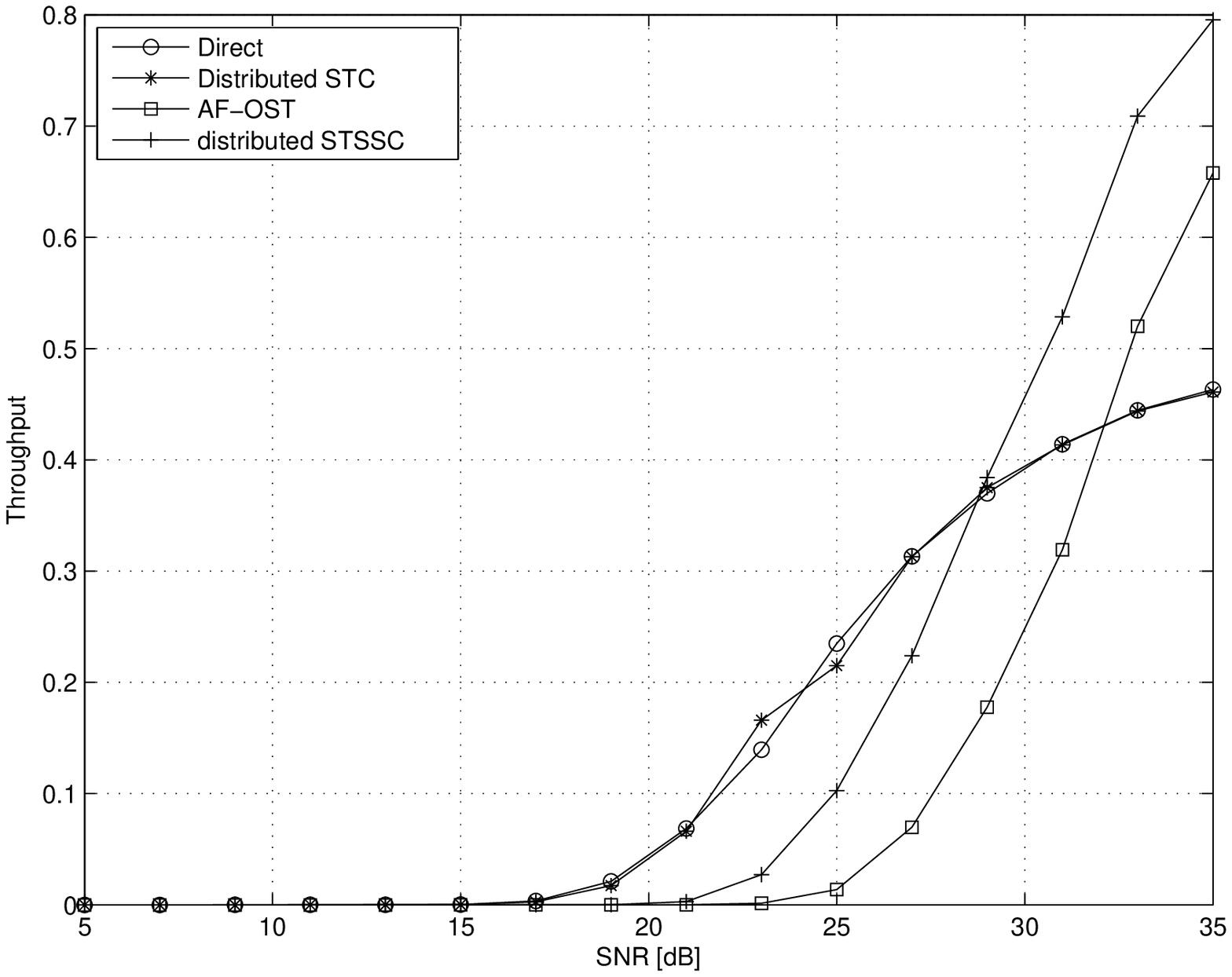}}

%\subfigure[QPSK]{ \includegraphics[keepaspectratio,width = 0.5\linewidth]{test_ber_2snd_qpsk.eps}}%\hspace{-0.2cm}
%\subfigure[QPSK]{ \includegraphics[keepaspectratio,width = 0.5\linewidth]{test_tput_2snd_qpsk.eps}}
\caption{BER and throughput results for two sources with Alamouti-type code with $T=2$ and all $|h_{x,y}|=1$.}
 \label{fig:results1}
\end{center}
\end{figure}

\subsection{Results for Two Sources}
The related results for two sources and two relays can be seen in Fig.~\ref{fig:results1}. For this case an Alamouti-type of code was employed by the relays. The $Direct$ mode corresponds to the performance of the point-to-point link. The first observation is that the \emph{Distributed STC} system performs similarly with a \emph{Direct} transmission. This is expected since the diversity gain that we obtain from the \emph{Distributed STC} in the $R\rightarrow D$ links, is minimized only because of decoding errors at the relay for the $S\rightarrow R$ transmissions. Now the $AF$-$OST$ scheme performs very well but in the high SNR regime. This is because of the noise amplification that occurs at the relays but when the signal quality is good, the joint ML decoding algorithm can have significant impact on throughput. On the other hand, $STSSC$ performs superior to $AF$-$OST$ even in the lower SNR regime because of the collected diversity benefits from the employed STC.

%For QPSK modulation the results can be seen in Fig.~\ref{fig:results2}.

\begin{figure}[t]
\begin{center}
\subfigure[BER]{ \includegraphics[keepaspectratio,width = 0.495\linewidth]{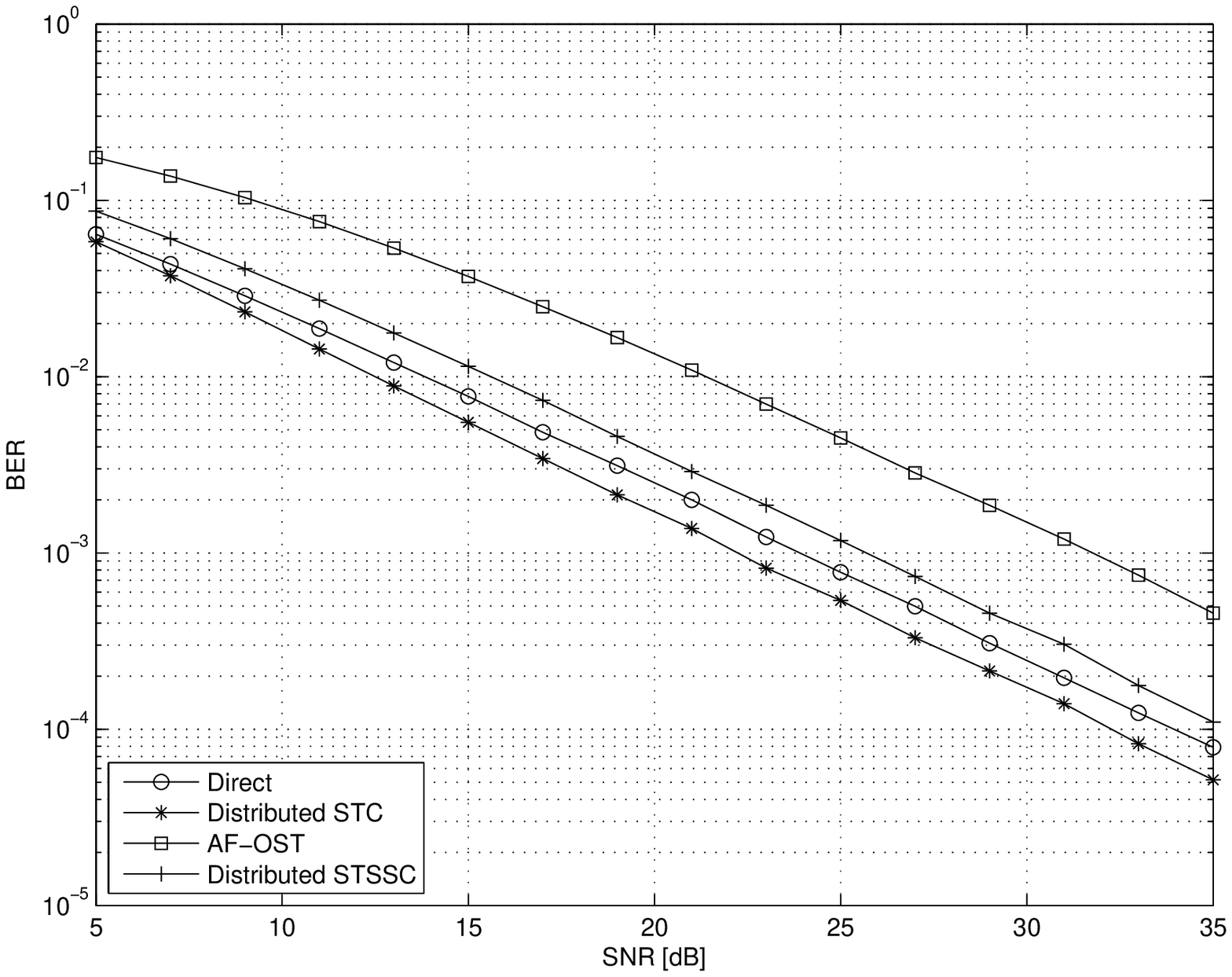}}\hspace{-0.3cm}
\subfigure[Throughput]{ \includegraphics[keepaspectratio,width = 0.495\linewidth]{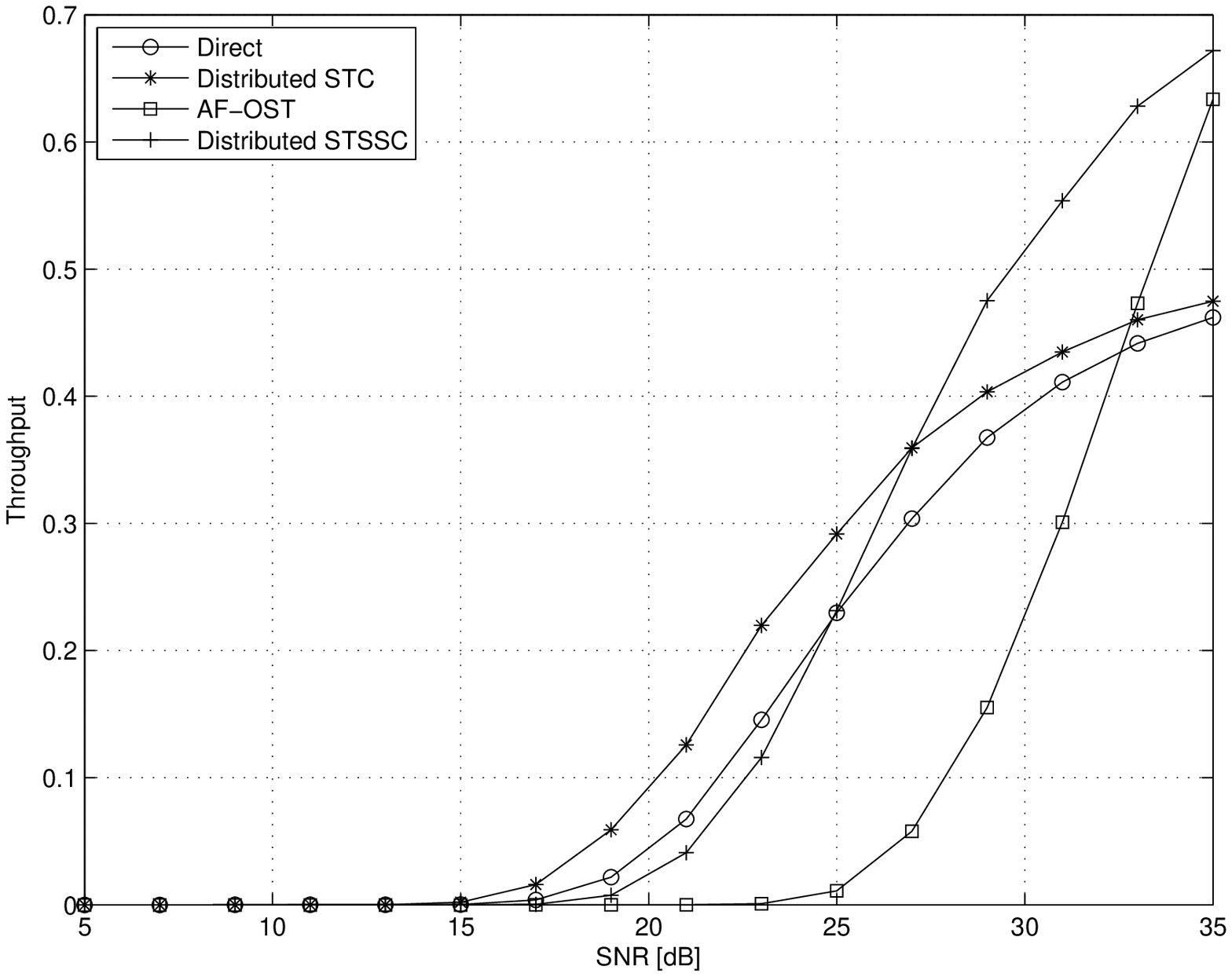}}
\caption{BER and throughput results for three sources with $C_{3,4}$ code, $T=4$, and all $|h_{x,y}|=1$.}
\label{fig:results3}
\end{center}
\end{figure}

\begin{figure}[t]
\begin{center}
\subfigure[BER]{ \includegraphics[keepaspectratio,width = 0.495\linewidth]{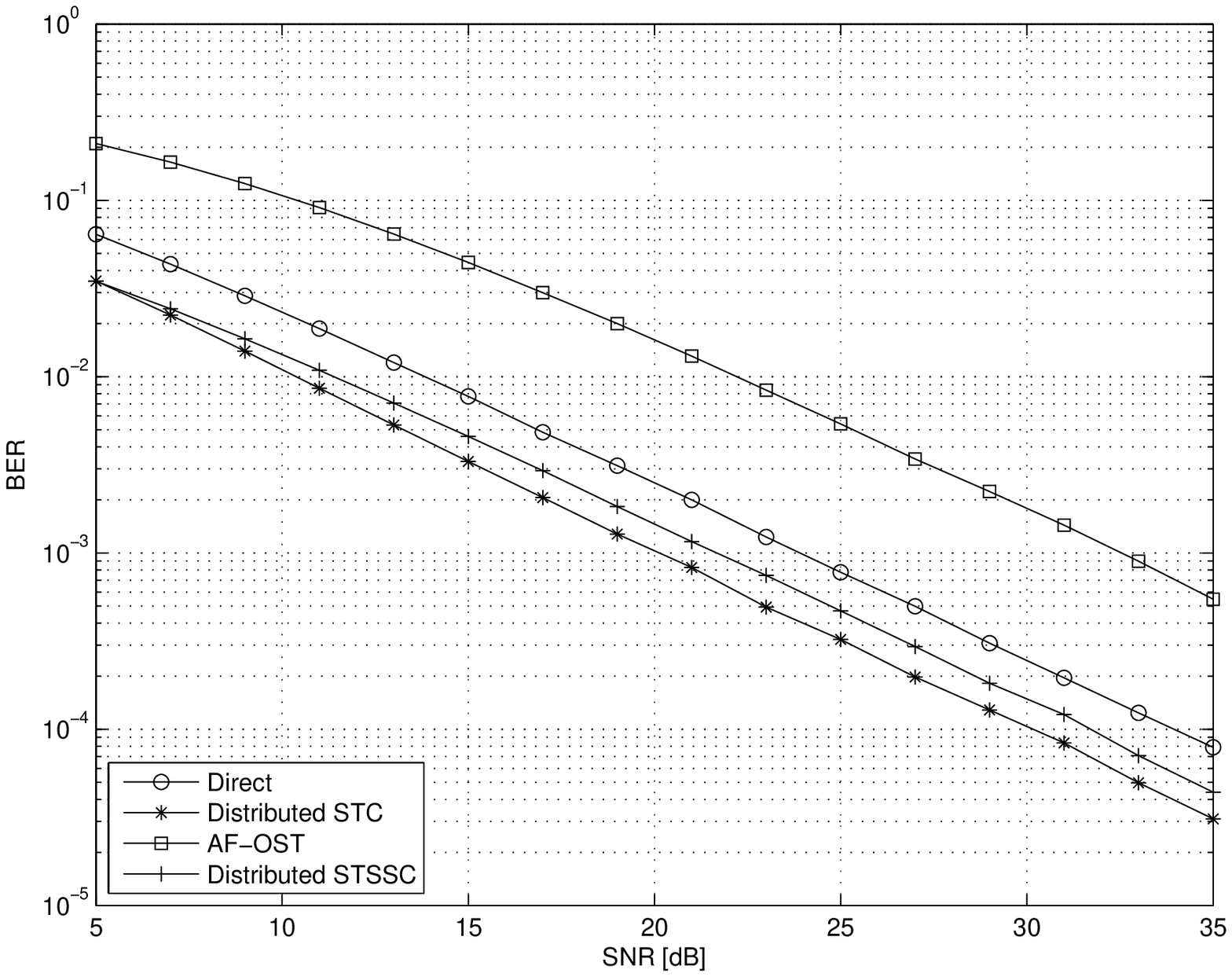}}\hspace{-0.3cm}
\subfigure[Thorughput]{ \includegraphics[keepaspectratio,width = 0.495\linewidth]{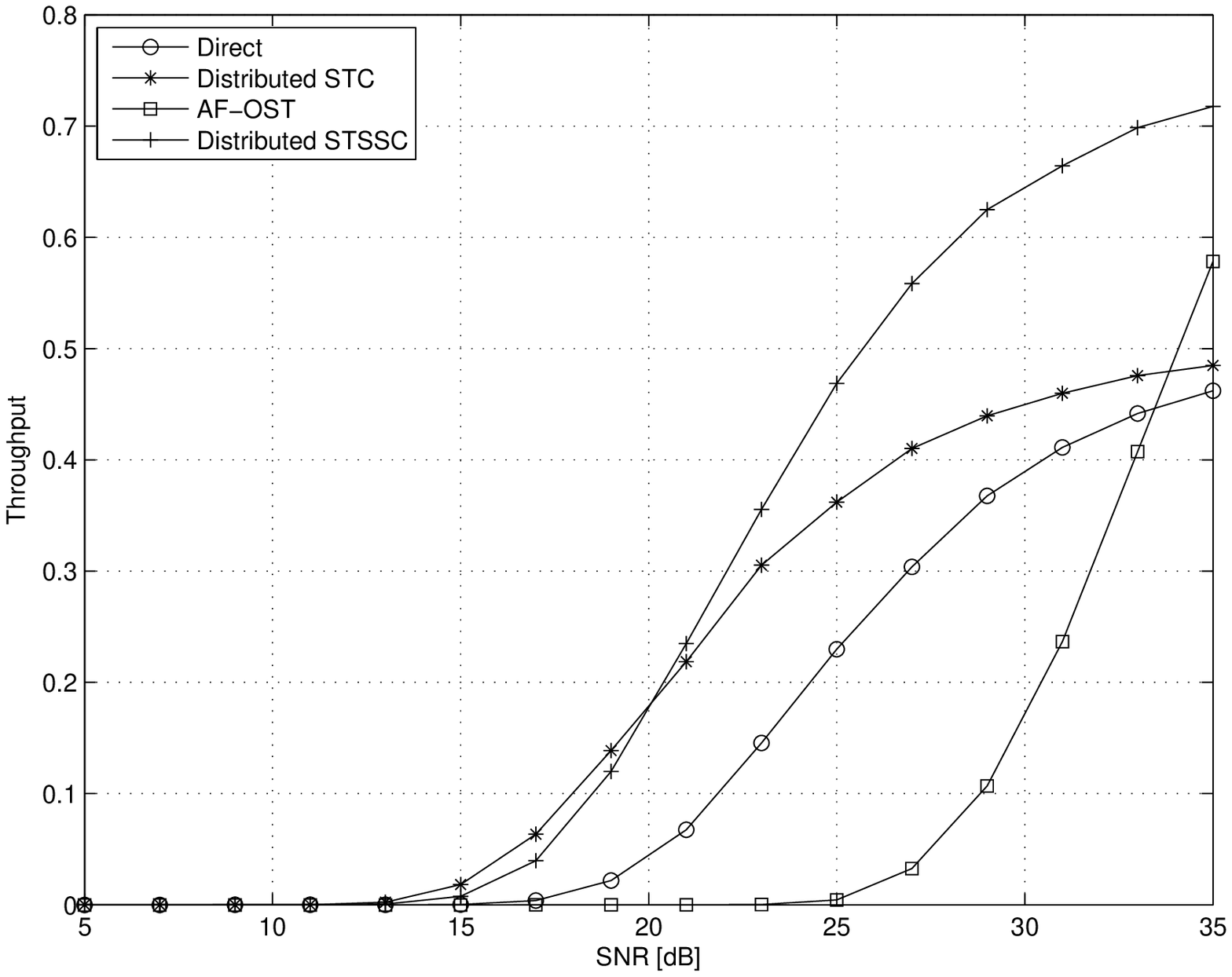}}
\caption{BER and throughput results for four sources with $C_{4,4}$ code, $T=4$, and all $|h_{x,y}|=1$.}
 \label{fig:results4}
\end{center}
\end{figure}

\subsection{Results for Three and Four Sources}
The selected code for three sources named $C_{3,4}$, is not a full-rate code but it is orthogonal while for four sources the code $C_{4,4}$ is a full-rate orthogonal code~\cite{book:coding-for-mimo}. The related results can be seen in Fig.~\ref{fig:results3} and Fig.~\ref{fig:results4}. When compared with the $N$=2 system, the case of $N$=3 with the \emph{Distributed STC} already performs better since another node is used for the STC. It now outperforms \emph{Direct} transmission even if the impact of decoding errors at the relay still exists. On the other hand $AF$-$OST$ presents minor performance decrease for $N$=3 and even more visible for $N$=4, which indeed shows that $N$=2 is the optimal choice for this protocol. This is primarily due to noise amplification at the relays that is increased as more of them are used. The more encouraging results can be seen for $STSSC$ where performance is significantly improved as the number of sources is increased while it also outperforms significantly the \emph{Distributed STC} even in the low SNR regime. This is an important result since at this point we can observe a reversal in the performance trend when compared to $AF$-$OST$, i.e. the BER is minimized when more sources are involved and superimpose their signals. The reason for this behavior is that the diversity benefits from the use of the STC increase significantly the decoding performance of the joint ML detector at the receiver. The key issue is that the same superimposed symbol, even though it receives diversity equal to the number of relays with $AF$-$OST$, it receives a diversity equal to the number of relays times $T$ since it is coded and re-forwarded from them at each symbol slot.

\section{Conclusions}
\label{section:conclusions}
In this paper we introduced the concept of space-time superimposed symbol coding. We showed that wireless signals that are superimposed over the air can still reap the benefits of space-time coding if nodes in the wireless network cooperate. This is accomplished first by applying the space-time code not on the symbols of interest themselves, but on the non-decodable superimposed signal of several symbols, and second by shifting the decoding to the final destinations. Thus, relays are still less complex in the sense that they are not required to perform any decoding operation besides the STC creation and at the same time their presence is fully exploited. Performance results show that significant throughput benefits can be observed over a distributed STC scheme. Also significant performance improvement is observed in the low SNR regime, an area where cooperative PLNC schemes traditionally suffer due to noise amplification.

\bibliography{../../tony-bib}
\end{document}